\documentclass[a4paper]{article}

\usepackage{INTERSPEECH2022}
\usepackage{subfigure}
\usepackage{multirow}
\usepackage{array}
\usepackage{color}
\newcolumntype{P}[1]{>{\centering\arraybackslash}p{#1}}
\usepackage{array}
\usepackage{booktabs}
\usepackage{array}
\usepackage{mdwmath}
\usepackage{mdwtab}
\usepackage{eqparbox}
\usepackage{url}
\usepackage{hyperref}
\usepackage{multirow}
\usepackage{makecell}
\usepackage{bbding}
\usepackage{subfigure}
\newcolumntype{P}[1]{>{\centering\arraybackslash}p{#1}}

\title{DDS: A new device-degraded speech dataset for speech enhancement}
\name{Haoyu Li$^1{^,}^2$, Junichi Yamagishi$^1{^,}^2$}
\address{
  $^1$National Institute of Informatics, Japan\\
  $^2$The Graduate University for Advanced Studies (SOKENDAI), Japan}
\email{haoyuli@nii.ac.jp, jyamagis@nii.ac.jp}

\begin{document}

\maketitle
\begin{abstract}
A large and growing amount of speech content in real-life scenarios is being recorded on consumer-grade devices in uncontrolled environments, resulting in degraded speech quality. 
Transforming such low-quality device-degraded speech into high-quality speech is a goal of speech enhancement (SE). This paper introduces a new speech dataset, DDS, to facilitate the research on SE. DDS provides aligned parallel recordings of high-quality speech (recorded in professional studios) and a number of versions of low-quality speech, producing approximately 2,000 hours speech data. 
The DDS dataset covers 27 realistic recording conditions by combining diverse acoustic environments and microphone devices, and each version of a condition consists of multiple recordings from six microphone positions to simulate different noise and reverberation levels.
We also test several SE baseline systems on the DDS dataset and show the impact of recording diversity on performance.

\end{abstract}

\section{Introduction}
\label{sec:intro}

High-quality speech is desired not only in speech communication systems such as mobile telephony but also for speech-generation tasks such as text-to-speech (TTS) \cite{ze2013statistical} and voice conversion (VC) \cite{chen2014voice}.
However, much of the speech content in real-life scenarios is recorded using non-professional devices (e.g., smartphones and laptops) in non-acoustically treated environments (e.g., homes and offices), where environmental noise, reverberation, and distortion of microphone frequency response degrade the quality of the speech. In this paper, we refer to speech that has been collected under such uncontrolled recording conditions as \textit{device-degraded} speech. 

Various speech enhancement (SE) techniques for raising the quality of device-degraded speech have been attracting attention.
Recently, deep neural network (DNN)-based SE methods \cite{lu2013speech, xu2014regression, weninger2015speech} have become mainstream and shown significant performance improvement over traditional methods \cite{boll1979suppression, nakatani2010speech}. Training a data-driven DNN model usually requires large datasets, but the existing datasets are relatively smaller compared to those used in other domains such as image classification \cite{deng2009imagenet}. Also, most datasets consist of only synthetic noisy speech rather than real noisy recordings. Although noisy speech can be obtained easily enough by adding clean speech with random noise segments \cite{reddy2019scalable, hu2007subjective} or convolving with room impulse responses \cite{valentini2016investigating}, Reddy \MakeLowercase{\textit{et al.}} \cite{reddy2020interspeech} pointed out that models trained on synthetic datasets often degrade significantly on real recordings.
This is mostly because the realistic device degradation cannot be perfectly simulated by synthetic datasets. For example, the measured transfer functions cannot capture the nonlinear reverberation and nonlinear distortion of microphone occurred real-world recording. 

In this work, we present a large-scale public dataset consisting of realistic device-degraded speech, named \textit{DDS}, to better facilitate the study of SE. DDS contains real recordings that are collected in diverse realistic environments using various microphone devices. 
More specifically, DDS is built on top of two existing datasets: DAPS \cite{mysore2014can} and VCTK \cite{yamagishi2019cstr}. We play clean speech recordings (four hours from DAPS and eight hours from VCTK) and re-record waveforms in nine environments (two offices, two conference rooms, three working studios\footnote{Specifically: a photo studio, a capture studio, and a voice studio.}, one living room, and one waiting room) on three different devices (one MEMS and two condenser microphones), producing 27 different recording conditions. 
For each condition, recordings are conducted with six microphone positions to simulate different noise and reverberation levels. In total, DDS contains 1,944 hours (3 devices $\times$ 9 environments $\times$ 6 positions $\times$ 12 hours) of realistic recordings. 
As far as we are concerned, this is the largest public dataset comprehensively covering various recording factors (i.e., environment, device, and position). In addition to the study of SE, it can be used in research domains such as domain adaptation in automatic speech recognition (ASR) \cite{sun2017unsupervised}, TTS/VC from found voice data \cite{yang2020adversarial}, and replay spoof detection in automatic speaker verification (ASV) \cite{todisco2019asvspoof}. The dataset is publicly available online: \url{https://doi.org/10.5281/zenodo.5464104}\footnote{We only released a down-sampled (16 kHz) version of DDS due to limited drive storage capacity. If the reader is interested in acquiring the full version, please contact haoyuli@nii.ac.jp and jyamagis@nii.ac.jp.}.

Section~\ref{sec:rw} of this paper reviews the relevant work.  Section~\ref{sec:overview} gives the details of the DDS dataset. In Section~\ref{sec:exp}, we test several SE baseline systems on the dataset and report the results. We conclude in Section~\ref{sec:conclu} with a brief summary.

\begin{figure}[t]
\begin{minipage}[b]{0.99\linewidth}
    \centering
    \subfigure[Schematic diagram of recording setup]
    {\includegraphics[height=5.7cm, width=7.7cm]{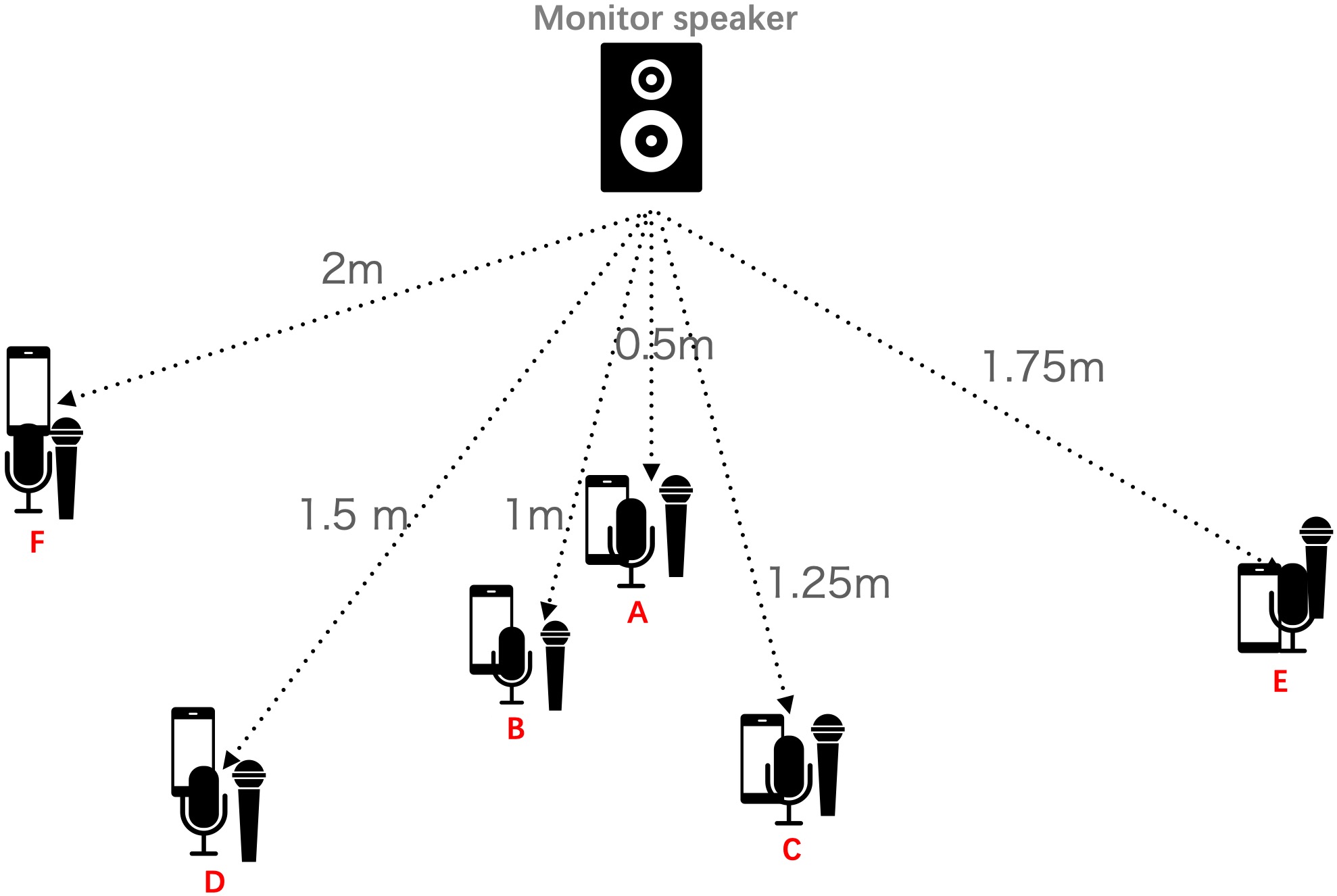}}
    \quad\quad
    \vspace{1.5mm}
    \subfigure[Example of device recording in the \textit{livingroom1}]
     {\includegraphics[height=5.7cm, width=7.7cm]{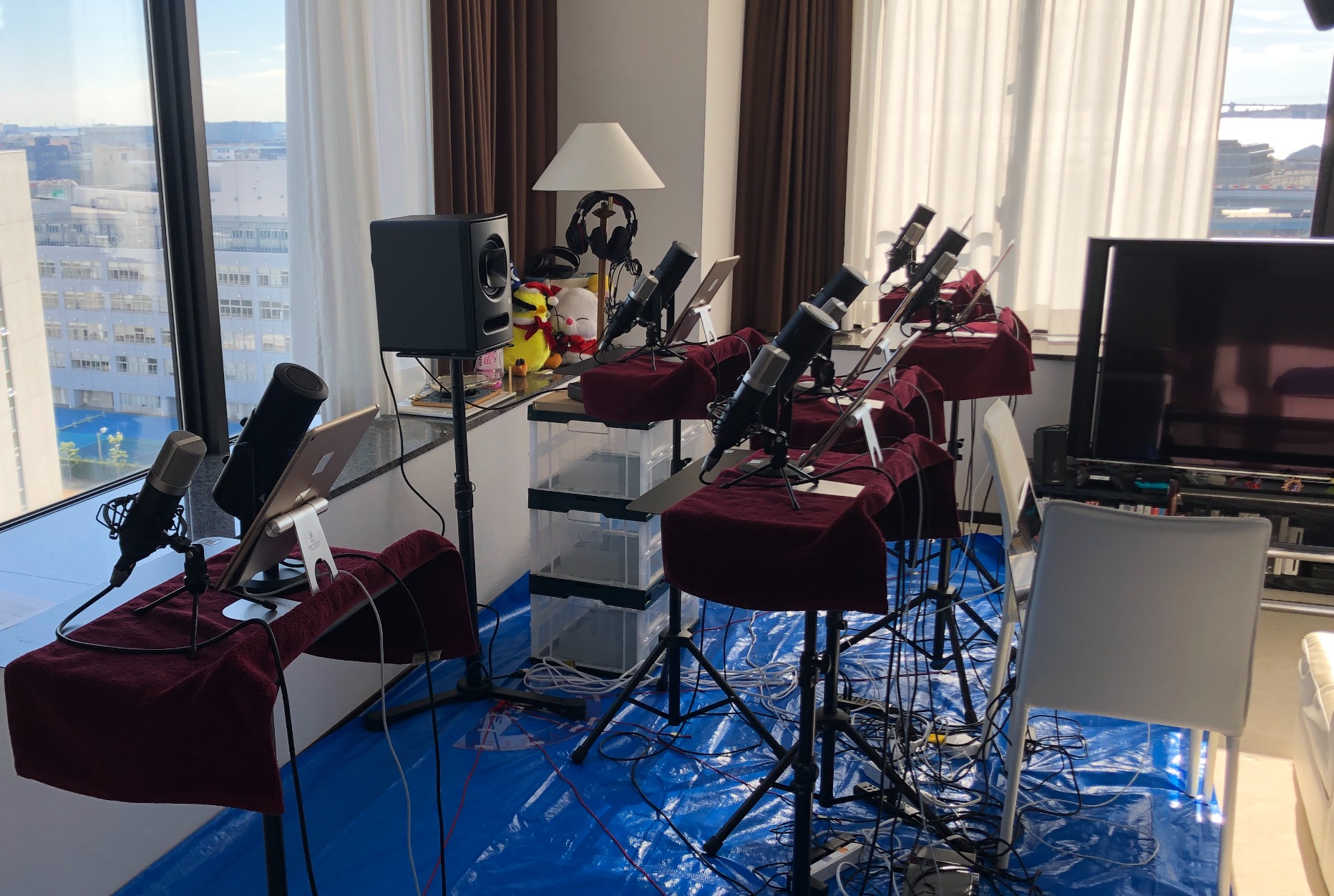}}
    \caption{Recording setup. Under each environment, studio-quality speech is played through a monitor loudspeaker and re-recorded on three devices (iPad Air, Uber Mic, and MPM-1000) at six (A\textendash E) positions.}
    \label{fig:setup}
\end{minipage}
\end{figure}

\section{Related Work}
\label{sec:rw}

Many speech datasets have been released \cite{reddy2019scalable, valentini2016investigating, mysore2014can} for the purpose of SE research. However, as mentioned in Section~\ref{sec:intro}, most of them contain only synthetic noisy speech (e.g., MS-SNSD \cite{reddy2019scalable} and Valentini's dataset \cite{valentini2016investigating}) and disregard microphone variability. Recently, Mathur \MakeLowercase{\textit{et al.}} released Libri-Adapt \cite{mathur2020libri}, which contains real recordings on six different microphones. However, as Libri-Adap is primarily developed for ASR research, the quality of its clean speech set (Librispeech-clean-100 dataset \cite{panayotov2015librispeech}) is not sufficient for tasks such as TTS. Also, the variability in acoustic environments is simply simulated by artificially adding different types of noises instead of recording in actual rooms. 

Our work is most closely related to the work on the DAPS dataset \cite{mysore2014can}, in which speech data are collected on different devices in realistic environments. Compared to their work, our dataset has a much larger size (DDS: 1,944 hours; DAPS: 50 hours) and contains more diverse recording conditions (DDS: 27 conditions; DAPS: 12 conditions). Furthermore, instead of recording speech at a certain fixed position, DDS dataset covers variability in the microphone positions.

\section{Dataset Overview}
\label{sec:overview}
In this section, we explain how we collected the DDS dataset and conduct an initial analysis. 
Table~\ref{tab:ovdataset} gives an overview of the dataset settings.

\begin{table}[t]
    \caption{Overview of dataset settings. \textit{MEMS} and \textit{condenser} denote microphone types. For device position, parameter \textit{(distance, angle)} denotes the distance and angle between device and sound source, respectively.}
    \vspace{1.5mm}
    \label{tab:ovdataset}
    \centering
    \renewcommand\arraystretch{1.3}
     \scalebox{0.93}{
    \begin{tabular}{c c c}
       
        \toprule
          \textbf{Setting} &
           \textbf{Count} &
           \textbf{Description} \\
          \hline
            \vspace{0.2mm}
          Speech materials & 2 & \makecell[c]{DAPS, VCTK clean sets} \\
          \hline
            \vspace{0.2mm}
          Environments & 9 & \makecell[c]{conference rooms (2), \\ offices (2), studios (3), \\living room (1), \\waiting room (1)}\\
          \hline
            \vspace{0.2mm}
          Devices & 3 & \makecell[c]{iPad Air (MEMS), \\ Uber Mic (condenser), \\ MPM-1000 (condenser)}\\
          \hline
            \vspace{0.2mm}
          Device positions & 6 & \makecell[c]{A(50 cm, 0\textdegree), B(100 cm, 15\textdegree) \\ 
          C(125 cm, 30\textdegree), D(150 cm, 45\textdegree) \\ 
          E(175 cm, 60\textdegree), F(200 cm, 75\textdegree)} \\

        \bottomrule
        
    \end{tabular}
    }
    \vspace{1mm}
\end{table}

\subsection{Speech materials}
\label{sec: dataset_overview}
Clean speech materials are selected from the DAPS \cite{mysore2014can} and VCTK \cite{yamagishi2019cstr} datasets, which both contain professional voice recordings. 
Specifically, the DAPS portion has four hours of speech data consisting of 20 speakers (ten female and ten male), and the VCTK portion has eight hours\footnote{We only selected part of VCTK speech instead of using the entire set.} of speech data consisting of 28 speakers (14 female and 14 male). As shown in Fig.~\ref{fig:setup}, we played and recorded speech using devices at a sampling rate of 48 kHz. To avoid the probable bias caused by the loudspeaker characteristics, we used a high-quality coaxial monitor speaker (Presonus Sceptre S6\footnote{https://www.presonus.com/products/sceptre-s6}) with very nice flat frequency response. For the DAPS portion, we re-sampled speech files into 44.1 kHz to match the original sampling rate of the DAPS clean set. Finally, we applied a cross-correlation algorithm to align the recorded speech with the original clean speech.

\subsection{Environments}
All recordings were conducted in realistic rooms\footnote{Details of room information (e.g., room size) and text scripts are included in the released DDS dataset.}. We selected a total of nine rooms with different layouts and sizes: two conference rooms, two offices, three studios, one living room, and one waiting room. Each room had a certain level of environmental noise and reverberation. It is worth noting that there is no constraint on the room noise. For example, the noise collected during recording may contain the sound of air conditioner, computer fans, or outdoor noise. Such background noise is close to that occurred in real-world recording, e.g., in home and office.

\subsection{Devices and recording positions}
Table~\ref{tab:ovdataset} lists the three microphone devices used during recording. These were a micro-electromechanical system (MEMS)-processed microphone, which is of small size and commonly embedded in smart devices, and two condenser microphones, which can offer a better sound quality than the MEMS microphones.

In addition to recording device, we conducted multiple recordings at six different positions for each device in each environment. 
The closest position was set to 50 cm directly in front of the speaker, while the farthest was set to 200 cm and at 75\textdegree  angle from the speaker. In this manner, we collected replayed speech with various noise and reverberation levels for each recording condition.

\subsection{Summary of DDS dataset}
In total, the DDS dataset consists of 9 environment settings and 3 device settings, resulting in a total of 27 recording conditions. Each condition consists of 83,058 speech files (13,843 files $\times$ 6 positions) at sampling rates of 44.1 kHz (for the DAPS portion) and 48 kHz (for the VCTK portion).

\subsection{Initial analysis of DDS}
\label{sec: analysis}
We conducted an analysis to investigate the effects of the various environments and devices on recording quality. We used PESQ \cite{rix2001perceptual} and ESTOI \cite{jensen2016algorithm} measures to evaluate objective speech quality and intelligibility, respectively. Tables~\ref{tab:env}, ~\ref{tab:device}, and~\ref{tab:pos} list the average scores under different conditions of environment, device, and position, respectively. We can clearly see that all recording factors dramatically affect speech quality and intelligibility. 
For example, as shown in Table~\ref{tab:env}, recording quality is directly related to room environment. Table~\ref{tab:device} shows that the condenser microphones (Uber Mic and MPM-1000) can offer a better sound quality than the MEMS one (iPad). Table~\ref{tab:pos} shows that speech recorded at a closer position has a better quality. In summary, these results demonstrate that DDS provides a sufficiently large variation of speech data to comprehensively cover common recording factors.

\begin{table}[t]
    \caption{Average PESQ and ESTOI scores in different environments.}
    \vspace{1mm}
    \label{tab:env}
    \centering
    \renewcommand\arraystretch{1.2}
    \scalebox{1.0}{
    \begin{tabular}{||c|c|c||c|c||}
        \hline
          \multirow{2}{*}{Environment} &
           \multicolumn{2}{c||}{DAPS portion \quad} & \multicolumn{2}{c||}{VCTK portion \quad}  \\
           \cline{2-5}
           & PESQ & ESTOI & PESQ & ESTOI\\
          \hline
          \hline
        confroom1 & 2.34 & 0.715 & 2.58 & 0.630 \\
        confroom2 & 1.98 & 0.617 & 2.27 & 0.527 \\
        office1 & 2.60 & 0.758 & 2.80 & 0.660 \\
        office2 & 2.31 & 0.724 & 2.54 & 0.627 \\
        studio1 & 2.37 & 0.725 & 2.59 & 0.602 \\
        studio2 & 3.01 & 0.815 & 3.10 & 0.735 \\
        studio3 & 3.10 & 0.811 & 3.16 & 0.735 \\
        waitingroom1 & 3.02 & 0.796 & 3.13 & 0.722 \\
        livingroom1 & 2.34 & 0.723 & 2.61 & 0.647 \\
        \hline
    \end{tabular}
    \vspace{1mm}
    }
\end{table}

\begin{table}[t]
    \caption{Average PESQ and ESTOI scores for different devices.}
    \vspace{1mm}
    \label{tab:device}
    \centering
    \renewcommand\arraystretch{1.2}
    \scalebox{1.03}{
    \begin{tabular}{||c|c|c||c|c||}
        \hline
          \multirow{2}{*}{Device} &
           \multicolumn{2}{c||}{DAPS portion \quad} & \multicolumn{2}{c||}{VCTK portion \quad}  \\
           \cline{2-5}
           & PESQ & ESTOI & PESQ & ESTOI\\
          \hline
          \hline
        iPad & 2.35 & 0.688 & 2.56 & 0.585 \\
        Uber Mic & 2.66 & 0.767 & 2.85 & 0.684 \\
        MPM-1000 & 2.68 & 0.773 & 2.86 & 0.693 \\
        \hline
    \end{tabular}
    \vspace{1mm}
    }
\end{table}

\begin{table}[t]
    \caption{Average PESQ and ESTOI scores with different device positions.}
    \vspace{1mm}
    \label{tab:pos}
    \centering
    \renewcommand\arraystretch{1.2}
    \scalebox{0.98}{
    \begin{tabular}{||c|c|c||c|c||}
        \hline
          \multirow{2}{*}{Device position} &
           \multicolumn{2}{c||}{DAPS portion \quad} & \multicolumn{2}{c||}{VCTK portion \quad}  \\
           \cline{2-5}
           & PESQ & ESTOI & PESQ & ESTOI\\
          \hline
          \hline
        A (50cm, 0\textdegree) & 3.22 & 0.901 & 3.32 & 0.840 \\
        B (100cm, 15\textdegree) & 2.77 & 0.810 & 2.94 & 0.728 \\
        C (125cm, 30\textdegree) & 2.57 & 0.770 & 2.78 & 0.680 \\
        D (150cm, 45\textdegree) & 2.44 & 0.720 & 2.65 & 0.624 \\
        E (175cm, 60\textdegree) & 2.27 & 0.656 & 2.50 & 0.557 \\
        F (200cm, 75\textdegree) & 2.11 & 0.597 & 2.35 & 0.495 \\
        \hline
    \end{tabular}
    \vspace{1mm}
    }
\end{table}

\section{Baseline Experiments}
\label{sec:exp}

\begin{figure*}[t]
\begin{minipage}[b]{\linewidth}
    \centering
    \subfigure[PESQ score]
    {\includegraphics[height=5.9cm, width=8.3cm]{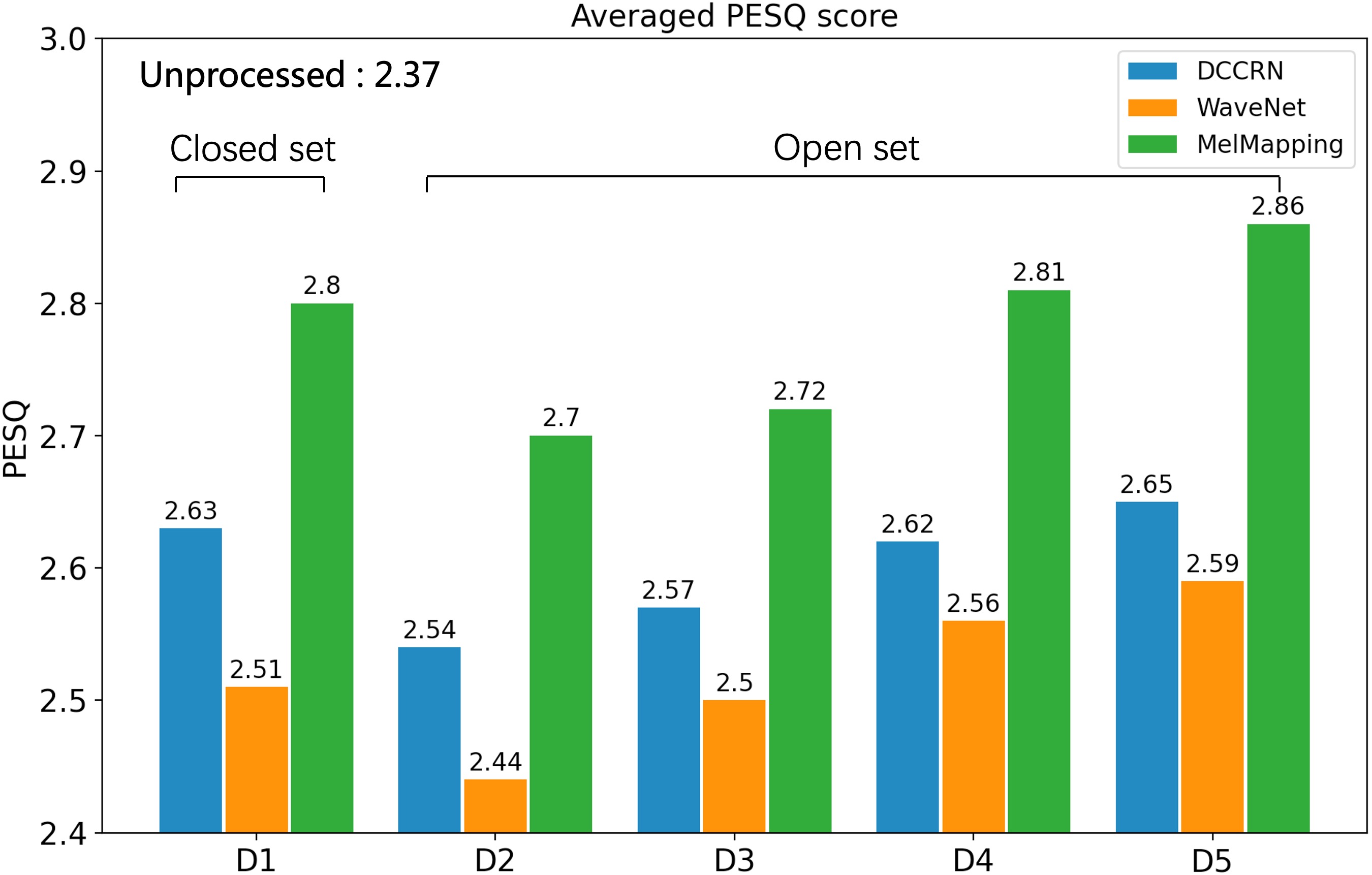}}
    \quad
    \subfigure[ESTOI score]
     {\includegraphics[height=5.9cm, width=8.3cm]{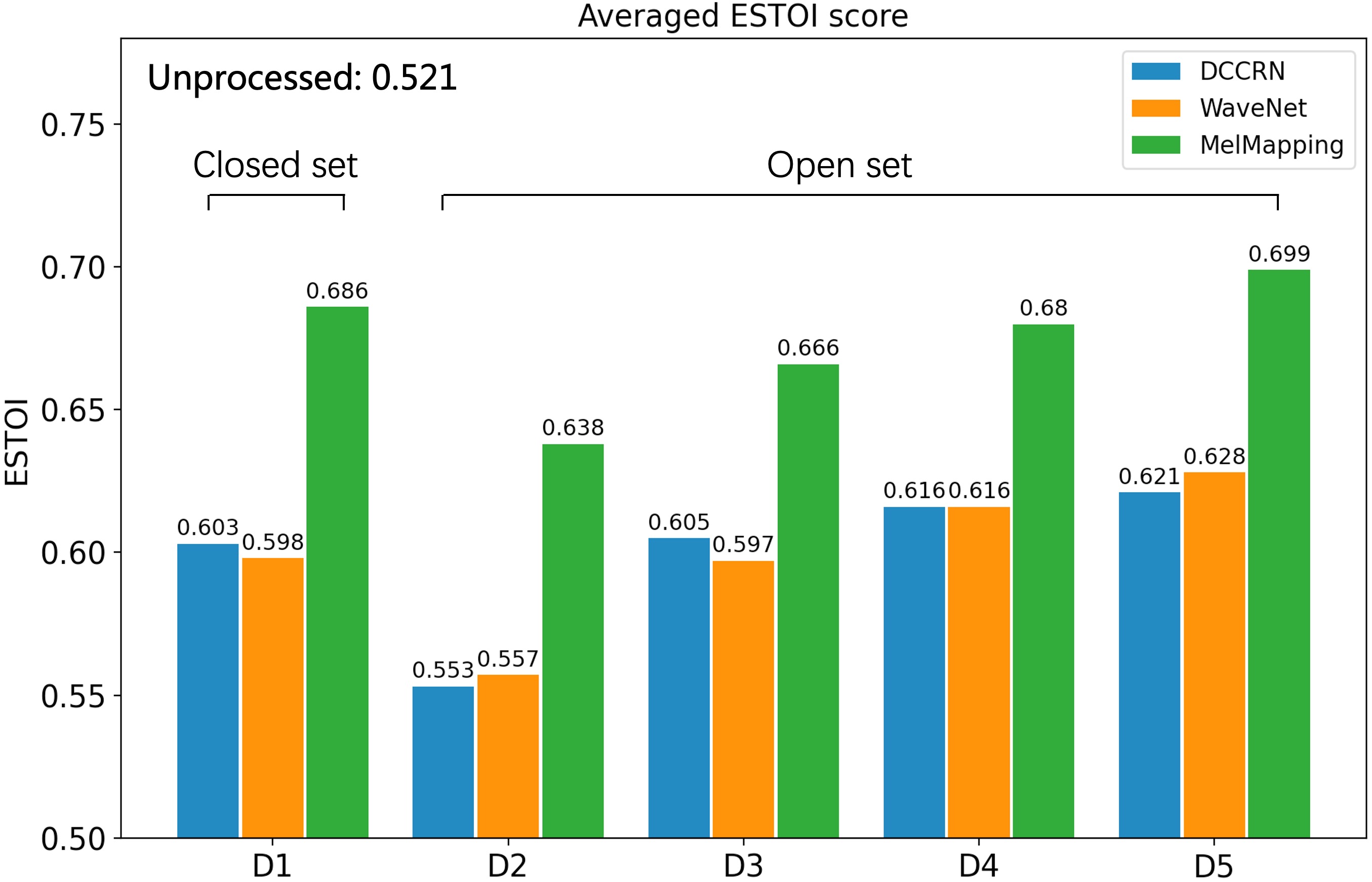}}
    \caption{Averaged PESQ and ESTOI scores for different speech enhancement systems on different training sets.}
    \label{fig:result}

\end{minipage}
\end{figure*}

In this section, we tested three baseline systems on the DDS dataset. We introduce and notate each system as follows.
\begin{itemize}
    \item \textbf{DCCRN}: Deep complex convolution recurrent network model \cite{hu2020dccrn} that performs speech denoising on complex-valued spectrogram instead of real-valued magnitude. This model showed good results in the Deep Noise Suppression (DNS) challenge 2020 \cite{reddy2020interspeech}. We reimplemented it using the official released codes\footnote{\url{https://github.com/huyanxin/DeepComplexCRN}} with the same loss function, i.e., scale-invariant signal-to-noise ratio (SI-SNR) \cite{le2019sdr}. The number of model parameters is 3.7M.
    \item \textbf{WaveNet}: An end-to-end waveform prediction model based on WaveNet \cite{su2019perceptually}, which is the backbone network architecture for many other waveform-domain enhancement models \cite{su2020hifi,su2021HSS}. We reimplemented it with the same model architecture and training objective (L1 loss on log spectrogram magnitude). The number of parameters is 8.4M.
    \item \textbf{MelMapping}: Our previously proposed system \cite{li2021enhancing} that predicts the clean Mel spectrogram and then reconstructs the speech signal using a universal WaveRNN vocoder \cite{kalchbrenner2018efficient}. The number of parameters is 11.7M.
\end{itemize}

We selected 500 utterances as the test set, and all these 500 utterances were recorded at F position (200 cm, 75\textdegree) in the \textit{livingroom1} by Uber Mic. To investigate the effect of recording diversity on performance, we selected and built five different training sub-sets from the original DDS dataset:
\begin{itemize}
    \item \textbf{D1:} Training utterances selected from \textit{\{Uber\}} device at A\textendash E positions in \textit{\{livingroom1\}} environment.
    \item \textbf{D2:} Training utterances selected from \textit{\{iPad\}} device at A\textendash E positions in \textit{\{confroom1\}} environment.
    \item \textbf{D3:} Training utterances selected from \textit{\{iPad, MPM-1000\}} devices at A\textendash E positions in \textit{\{confroom1\}} environment.
    \item \textbf{D4:} Training utterances selected from \textit{\{iPad, MPM-1000\}} devices at A\textendash E positions in \textit{\{confroom1, confroom2, office1, office2\}} environments.
    \item \textbf{D5:} Training utterances selected from \textit{\{iPad, MPM-1000\}} devices at A\textendash E positions in \textit{\{confroom1, confroom2, office1, office2, studio1, studio2, studio3, waitingroom1\}} environments.
\end{itemize}
Each training sub-set contains 40,000 (\# sentences $\times$ \# devices $\times$ \# environments $\times$ \# positions) training utterances of the same 32-hour duration.
\textbf{D1} shares the same recording condition (i.e., in living room by Uber Mic) as the test set, so we regard it as a closed-set task. For the remaining four training sets, they are unseen to the test set in terms of recording device and environment and regarded as open-set tasks. Also, they were intentionally designed with increasing recording diversity. For example, \textbf{D2} comprises only a single condition with one device and one environment, whereas \textbf{D5} comprises various conditions including two devices and eight environments. We used PESQ and ESTOI scores to evaluate the speech quality and intelligibility, respectively.

The results are plotted in Fig.~\ref{fig:result}. Among the compared systems, \textbf{MelMapping} performed best in terms of PESQ and ESTOI. Although \textbf{DCCRN} and \textbf{WaveNet} suppressed the additive noise well, they were less good at addressing the reverberation and device distortion, resulting in relatively lower scores. 
Besides, as mentioned in Section~\ref{sec: dataset_overview}, the alignment between device-degraded speech and its clean counterpart was done using cross-correlation algorithm, which inevitably results in a small time shift. When using time-domain loss functions, even a small time shift can degrade the performance.
This explains why the state-of-the-art \textbf{DCCRN}, which was trained on time-domain SI-SNR loss, performed not good in our experiments
The performance of each system also varied with the training set types. For all three systems, the performance on closed-set \textbf{D1} was significantly better than that on open-set \textbf{D2}, which indicates that the domain mismatch between training and test affects the enhancement performance significantly.
By increasing the recording diversity (from \textbf{D2} to \textbf{D5}), however, the performance can clearly be improved.
It is worth noting that all three systems achieved better or comparable results on \textbf{D5} (open-set but with the most diversity in recording devices and environments) than on closed-set \textbf{D1}. This further indicates that incorporating various recording conditions into training can help SE models generalize to the noisy recordings encountered in real-world scenarios. 
\vspace{-1mm}
\section{Conclusion}
\label{sec:conclu}
\vspace{-1mm}
In this paper, we introduced a large device-degraded speech dataset called DDS to facilitate the research on speech enhancement, especially the enhancement of real-world consumer-grade recordings. This dataset contains studio-quality clean speech and corresponding low-quality versions, with 1,944 hours of real recordings collected under 27 realistic conditions spanning three microphones and nine acoustic environments. We reported several baseline results on this dataset, and showed the beneficial effect of recording diversity of training data on model performance. The DDS dataset is publicly available at \url{https://zenodo.org/record/5464104}.

\section{Acknowledgements}

This work was supported in part by JST CREST VoicePersonae project under Grant JPMJCR18A6, Japan, in part by MEXT KAKENHI Grants, 18H04112 and 21H04906, Japan, and in part by SOKENDAI (The Graduate University for Advanced Studies), Japan.

\bibliography{template.bbl}

\begin{thebibliography}{10}
\providecommand{\url}[1]{#1}
\csname url@samestyle\endcsname
\providecommand{\newblock}{\relax}
\providecommand{\bibinfo}[2]{#2}
\providecommand{\BIBentrySTDinterwordspacing}{\spaceskip=0pt\relax}
\providecommand{\BIBentryALTinterwordstretchfactor}{4}
\providecommand{\BIBentryALTinterwordspacing}{\spaceskip=\fontdimen2\font plus
\BIBentryALTinterwordstretchfactor\fontdimen3\font minus
  \fontdimen4\font\relax}
\providecommand{\BIBforeignlanguage}[2]{{%
\expandafter\ifx\csname l@#1\endcsname\relax
\typeout{** WARNING: IEEEtran.bst: No hyphenation pattern has been}%
\typeout{** loaded for the language `#1'. Using the pattern for}%
\typeout{** the default language instead.}%
\else
\language=\csname l@#1\endcsname
\fi
#2}}
\providecommand{\BIBdecl}{\relax}
\BIBdecl

\bibitem{ze2013statistical}
H.~Zen, A.~Senior, and M.~Schuster, ``Statistical parametric speech synthesis
  using deep neural networks,'' in \emph{2013 IEEE International Conference on
  Acoustics, Speech and Signal Processing (ICASSP)}.\hskip 1em plus 0.5em minus
  0.4em\relax IEEE, 2013, pp. 7962--7966.

\bibitem{chen2014voice}
L.-H. Chen, Z.-H. Ling, L.-J. Liu, and L.-R. Dai, ``Voice conversion using deep
  neural networks with layer-wise generative training,'' \emph{IEEE/ACM
  Transactions on Audio, Speech, and Language Processing}, vol.~22, no.~12, pp.
  1859--1872, 2014.

\bibitem{lu2013speech}
X.~Lu, Y.~Tsao, S.~Matsuda, and C.~Hori, ``Speech enhancement based on deep
  denoising autoencoder.'' in \emph{Interspeech}, vol. 2013, 2013, pp.
  436--440.

\bibitem{xu2014regression}
Y.~Xu, J.~Du, L.-R. Dai, and C.-H. Lee, ``A regression approach to speech
  enhancement based on deep neural networks,'' \emph{IEEE/ACM Transactions on
  Audio, Speech, and Language Processing}, vol.~23, no.~1, pp. 7--19, 2014.

\bibitem{weninger2015speech}
F.~Weninger, H.~Erdogan, S.~Watanabe, E.~Vincent, J.~Le~Roux, J.~R. Hershey,
  and B.~Schuller, ``{Speech enhancement with {LSTM} recurrent neural networks
  and its application to noise-robust ASR},'' in \emph{International Conference
  on Latent Variable Analysis and Signal Separation}.\hskip 1em plus 0.5em
  minus 0.4em\relax Springer, 2015, pp. 91--99.

\bibitem{boll1979suppression}
S.~Boll, ``Suppression of acoustic noise in speech using spectral
  subtraction,'' \emph{IEEE Transactions on Acoustics, Speech, and Signal
  Processing}, vol.~27, no.~2, pp. 113--120, 1979.

\bibitem{nakatani2010speech}
T.~Nakatani, T.~Yoshioka, K.~Kinoshita, M.~Miyoshi, and B.-H. Juang, ``Speech
  dereverberation based on variance-normalized delayed linear prediction,''
  \emph{IEEE Transactions on Audio, Speech, and Language Processing}, vol.~18,
  no.~7, pp. 1717--1731, 2010.

\bibitem{deng2009imagenet}
J.~Deng, W.~Dong, R.~Socher, L.-J. Li, K.~Li, and L.~Fei-Fei, ``{ImageNet: A
  large-scale hierarchical image database},'' in \emph{2009 IEEE conference on
  computer vision and pattern recognition}.\hskip 1em plus 0.5em minus
  0.4em\relax IEEE, 2009, pp. 248--255.

\bibitem{reddy2019scalable}
C.~K. Reddy, E.~Beyrami, J.~Pool, R.~Cutler, S.~Srinivasan, and J.~Gehrke, ``A
  scalable noisy speech dataset and online subjective test framework,''
  \emph{arXiv preprint arXiv:1909.08050}, 2019.

\bibitem{hu2007subjective}
Y.~Hu, ``Subjective evaluation and comparison of speech enhancement
  algorithms,'' \emph{Speech Communication}, vol.~49, pp. 588--601, 2007.

\bibitem{valentini2016investigating}
C.~Valentini-Botinhao, X.~Wang, S.~Takaki, and J.~Yamagishi, ``{Investigating
  RNN-based speech enhancement methods for noise-robust Text-to-Speech},'' in
  \emph{SSW}, 2016, pp. 146--152.

\bibitem{reddy2020interspeech}
C.~K. Reddy, V.~Gopal, R.~Cutler, E.~Beyrami, R.~Cheng, H.~Dubey,
  S.~Matusevych, R.~Aichner, A.~Aazami, S.~Braun \emph{et~al.}, ``{The
  Interspeech 2020 deep noise suppression challenge: Datasets, subjective
  testing framework, and challenge results},'' \emph{arXiv preprint
  arXiv:2005.13981}, 2020.

\bibitem{mysore2014can}
G.~J. Mysore, ``Can we automatically transform speech recorded on common
  consumer devices in real-world environments into professional production
  quality speech?—a dataset, insights, and challenges,'' \emph{IEEE Signal
  Processing Letters}, vol.~22, no.~8, pp. 1006--1010, 2014.

\bibitem{yamagishi2019cstr}
J.~Yamagishi, C.~Veaux, and K.~MacDonald, ``{CSTR VCTK Corpus}: English
  multi-speaker corpus for {CSTR Voice Cloning Toolkit} (version 0.92),'' 2019.

\bibitem{sun2017unsupervised}
S.~Sun, B.~Zhang, L.~Xie, and Y.~Zhang, ``An unsupervised deep domain
  adaptation approach for robust speech recognition,'' \emph{Neurocomputing},
  vol. 257, pp. 79--87, 2017.

\bibitem{yang2020adversarial}
S.~Yang, Y.~Wang, and L.~Xie, ``Adversarial feature learning and unsupervised
  clustering based speech synthesis for found data with acoustic and textual
  noise,'' \emph{IEEE Signal Processing Letters}, vol.~27, pp. 1730--1734,
  2020.

\bibitem{todisco2019asvspoof}
M.~Todisco, X.~Wang, V.~Vestman, M.~Sahidullah, H.~Delgado, A.~Nautsch,
  J.~Yamagishi, N.~Evans, T.~Kinnunen, and K.~A. Lee, ``{ASVspoof 2019: Future
  horizons in spoofed and fake audio detection},'' \emph{arXiv preprint
  arXiv:1904.05441}, 2019.

\bibitem{mathur2020libri}
A.~Mathur, F.~Kawsar, N.~Berthouze, and N.~D. Lane, ``{Libri-Adapt}: a new
  speech dataset for unsupervised domain adaptation,'' in \emph{2020 IEEE
  International Conference on Acoustics, Speech and Signal Processing
  (ICASSP)}.\hskip 1em plus 0.5em minus 0.4em\relax IEEE, 2020, pp. 7439--7443.

\bibitem{panayotov2015librispeech}
V.~Panayotov, G.~Chen, D.~Povey, and S.~Khudanpur, ``Librispeech: an {ASR}
  corpus based on public domain audio books,'' in \emph{2015 IEEE International
  Conference on Acoustics, Speech and Signal Processing (ICASSP)}.\hskip 1em
  plus 0.5em minus 0.4em\relax IEEE, 2015, pp. 5206--5210.

\bibitem{rix2001perceptual}
A.~W. Rix, J.~G. Beerends, M.~P. Hollier, and A.~P. Hekstra, ``{Perceptual
  evaluation of speech quality (PESQ)-a new method for speech quality
  assessment of telephone networks and codecs},'' in \emph{2001 IEEE
  International Conference on Acoustics, Speech and Signal Processing
  (ICASSP)}, vol.~2.\hskip 1em plus 0.5em minus 0.4em\relax IEEE, 2001, pp.
  749--752.

\bibitem{jensen2016algorithm}
J.~Jensen and C.~H. Taal, ``An algorithm for predicting the intelligibility of
  speech masked by modulated noise maskers,'' \emph{IEEE/ACM Transactions on
  Audio, Speech, and Language Processing}, vol.~24, no.~11, pp. 2009--2022,
  2016.

\bibitem{hu2020dccrn}
Y.~Hu, Y.~Liu, S.~Lv, M.~Xing, S.~Zhang, Y.~Fu, J.~Wu, B.~Zhang, and L.~Xie,
  ``{DCCRN: Deep complex convolution recurrent network for phase-aware speech
  enhancement},'' \emph{arXiv preprint arXiv:2008.00264}, 2020.

\bibitem{le2019sdr}
J.~Le~Roux, S.~Wisdom, H.~Erdogan, and J.~R. Hershey, ``{SDR--half-baked or
  well done?}'' in \emph{2019 IEEE International Conference on Acoustics,
  Speech and Signal Processing (ICASSP)}.\hskip 1em plus 0.5em minus
  0.4em\relax IEEE, 2019, pp. 626--630.

\bibitem{su2019perceptually}
J.~Su, A.~Finkelstein, and Z.~Jin, ``Perceptually-motivated
  environment-specific speech enhancement,'' in \emph{2019 IEEE International
  Conference on Acoustics, Speech and Signal Processing (ICASSP)}.\hskip 1em
  plus 0.5em minus 0.4em\relax IEEE, 2019, pp. 7015--7019.

\bibitem{su2020hifi}
J.~Su, Z.~Jin, and A.~Finkelstein, ``{HiFi-GAN: High-fidelity denoising and
  dereverberation based on speech deep features in adversarial networks},''
  \emph{arXiv preprint arXiv:2006.05694}, 2020.

\bibitem{su2021HSS}
------, ``{HiFi}-{GAN}-2: Studio-quality speech enhancement via generative
  adversarial networks conditioned on acoustic features,'' in \emph{WASPAA
  2021}, 2021.

\bibitem{li2021enhancing}
H.~Li, Y.~Ai, and J.~Yamagishi, ``Enhancing low-quality voice recordings using
  disentangled channel factor and neural waveform model,'' in \emph{2021 IEEE
  Spoken Language Technology Workshop (SLT)}.\hskip 1em plus 0.5em minus
  0.4em\relax IEEE, 2021, pp. 734--741.

\bibitem{kalchbrenner2018efficient}
N.~Kalchbrenner, E.~Elsen, K.~Simonyan, S.~Noury, N.~Casagrande, E.~Lockhart,
  F.~Stimberg, A.~Oord, S.~Dieleman, and K.~Kavukcuoglu, ``Efficient neural
  audio synthesis,'' in \emph{International Conference on Machine
  Learning}.\hskip 1em plus 0.5em minus 0.4em\relax PMLR, 2018, pp. 2410--2419.

\end{thebibliography}

\end{document}